# Effective potential engineering by emergent anisotropy in a tunable open-access microcavity

Yiming Li,[1] Xiaoxuan Luo,[1] Yaxin Guo,[1] Jiahuan Ren,[2,5] Teng Long,[2] Bohao Wang,[4] Yin Cai,[1] Chaowei Guo,[3] Yuanbin Qin,[3] Hongbing Fu,[2] Yanpeng Zhang,[1] Feng Yun,[1,4] Qing Liao,[2,a)] and Feng Li,[1,4,a)]

**AFFILIATIONS**

[1]Key Laboratory for Physical Electronics and Devices of the Ministry of Education & Shaanxi Key Lab of Information Photonic Technique, School of Electronic Science and Engineering, Faculty of Electronic and Information Engineering, Xi'an Jiaotong University, Xi'an 710049, China

[2]Beijing Key Laboratory for Optical Materials and Photonic Devices, Department of Chemistry, Capital Normal University, Beijing 100048, China

[3]Center for Advancing Materials Performance from the Nanoscale (CAMP-Nano), State Key Laboratory for Mechanical Behavior of Materials, Xi'an Jiaotong University, Xi'an 710049, China

[4]Solid-State Lighting Engineering Research Center, Xi'an Jiaotong University, Xi'an 710049, China

[5]School of Physical Science and Technology, Hebei University, Hebei Baoding 071000, China

[a)]**Authors to whom correspondence should be addressed:** liaoqing@cnu.edu.cn and felix831204@xjtu.edu.cn

**ABSTRACT**

Photonic spin-orbit (SO) coupling is an important physical mechanism leading to numerous interesting phenomena in the systems of microcavity photons and exciton-polaritons. We report the effect of SO coupling in a tunable open-access microcavity embedded with anisotropic active media. The SO coupling associated with the TE-TM splitting results in an emergent anisotropy, which further leads to fine energy splittings allowing clear observation of the full set of eigenstates, in sharp contrast with the isotropic situation which leads to the isotropic eigenstates of spin vortices. We show that the photonic potential can be engineered by playing with the relation between the emergent anisotropy and the cavity ellipticity. All the experimental results are well reproduced by the degenerate perturbation theory. Our results constitute a significant extension to the research field of microcavity spinoptronics, with potential applications in polarization control and optical property measurement of photonic devices and materials.

## I. INTRODUCTION

Spin-orbit (SO) coupling of photons and exciton-polaritons in Fabry-Perrot (FP) microcavities has brought rich physics and potential applications during the last two decades, leading to the research field of microcavity spinoptronics.[1-5] The transvers-electric-transvers-magnetic (TE-TM) splitting induced by the cavity mirrors serves as an effective gauge field affecting the pseudospin of cavity photons according to their orbital motion.[6] A series of significant phenomena have been demonstrated based on such mechanism, including the optical spin hall effect,[6-8] dark half solitons,[9,10] spin vortices in polariton quantum fluids[11] and polaritonic topological insulators.[12,13] Recently, more physical mechanisms have been applied to construct novel types of effective gauge field, by exerting magnetic field[14,15] and using anisotropic materials.[16,17] The latter, which induces an effect of emergent optical activity (OA), has led to Rashba-Dresselhaus SO coupling,[16,18,19] nontrivial topological valleys,[14,15,20,21] Voigt exceptional points,[22] divergent quantum metric,[23] and helical polariton lasing.[21]

In addition to free photonic and polaritonic gas in FP microcavities, the SO coupling effects in confined systems, i.e., microcavities with an artificial photonic potential, has also been intensively studied. The investigations reveal a series of interesting vector eigenstates in laterally confined microcavities, such as the spin vortices and antivortices experimentally demonstrated in an open-access microcavity[24] as well as a benzene-like photonic molecule.[25] Recently, it has been theoretically proved that the TE-TM splitting in a symmetric circular photonic potential leads to various vector eigenstates up to the order of infinite, including solid and hollow optical skyrmions, the distribution of which obeys a series of general rules.[26] Non-Hermitian manipulation of such cavity optical skyrmions is also proposed with the presence of exceptional points.[27] However, the TE-TM splitting alone does not fully lift the degeneracy of the higher order manifolds in symmetric confinements, leaving most of the energy levels still double degenerate.[26] As those studies were carried out with isotropic light-emitting materials, it is essential to further investigate the vector eigenstates in confined microcavity systems with anisotropic active materials, and reveal the different role that the TE-TM SO coupling would play compared to the isotropic counterparts, which would be associated with unique underlying physics and potential applications.

In this Article, we investigate the vector eigenstates in an anisotropic circular confinement potential under the effect of TE-TM splitting, with an experimental platform of a concave-planar open-access microcavity embedded with anisotropic organic microcrystals. Interestingly, the TE-TM splitting, being an isotropic effect, leads to emergent anisotropy which results in an effective ellipticity of the circular photonic potential. The existence of the effective ellipticity enables the observation of a complete lift of degeneracy, based on which we observe four orthogonally distributed eigenstates, between which the energy splittings can be engineered by the alignment of the material optical axis with respect to the cavity geometry. Our results reveal the unique role of SO coupling in anisotropically confined systems and therefore constitute a significant extension to the research field of microcavity spinoptronics, with potential applications in polarization control of lasers, building photonic simulators and measurements of unknown optical material properties.

## II. THE MICROCAVITY SYSTEM

The experimental platform is established by filling anisotropic organic microcrystals in an open-access microcavity, as sketched in Fig.1(a). The open-access microcavity consists of a bottom planar distributed Bragg reflector (DBR) and a top concave DBR individually controlled by nanopositioners,[28,29] in which the concave shape is fabricated by focused ion beam (FIB) milling.[30] The organic microcrystals, α-perylene prepared by space-confined self-assembly method,[20,31] are transferred onto the bottom DBR before the open-access microcavity is set up using nanopositioners. The microscopic image in Fig. 1(b) shows that the α-perylene microcrystals are square-shaped, with the diagonal directions (noted as x and y) being the directions of polarization of the ordinary light (o-light) and the extraordinary light (e-light).[20] The microcavity is characterized by real-space polarization-resolved micro-photoluminescence (μ-PL) at room temperature. More details about the sample fabrication and characterization are included in Materials and Methods.

The measured μ-PL spectra polarized along x and y directions, shown in Figs. 1(c) and 1(e) respectively, indicates that there exist two sets of orthogonally polarized modes corresponding to the o-light and the e-light, whilst each polarization contains several groups of peaks corresponding to different longitudinal orders. The enlarged spectra in Figs. 1(d) and 1(f) demonstrate a high quality factor (Q-factor) of ~7000 and ~6000, higher than most of the organic microcavities typically reported in literature.[32-35] To further identify the transvers modes, we obtain the real-space images measured with x and y linear polarizations, as presented in Fig. 1(g) and 1(h), by opening the entrance slit of the spectrometer. Several groups of Laguerre-Gauss (LG) modes are seen in each polarization set, similar as reported in literature.[28] The difference in photon energy (wavelength) of the same LG modes between the two orthogonal polarizations is a typical result of the linear birefringence of the organic microcrystal.

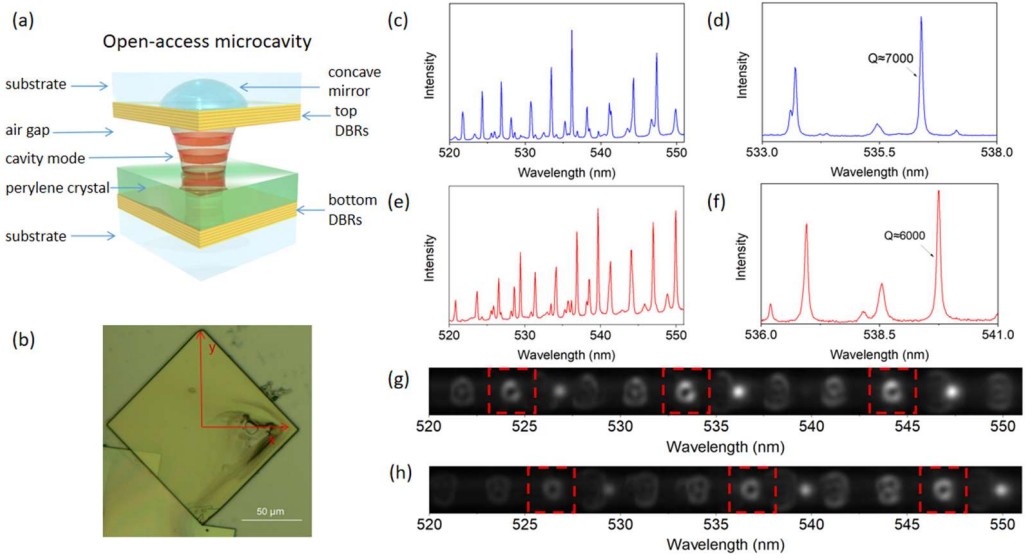

Figure 1: The sample and the μ-PL measurements. (a) The illustrative picture of the open-access microcavity. (b) The microscopic image of the perylene, while x and y axes correspond to the polarization directions of the e and o lights, respectively. (c) and (e) The polarization-resolved μ-PL spectra measured in the polarization of x (c) and y (e) directions. (d) and (f) The zoomed-in views of the spectra in (c) and (e). (g) and (h) The real-space μ-PL images taken by opening the spectrometer entrance slit, showing LG-like spatial distributions. The $LG_{0\pm1}$ modes are highlighted by dashed red squares.

### III. THEORETICAL MODEL AND ANALYSIS

As a convention, the LG modes are noted as $LG_{pl}$ where $p$ and $l$ are the radial and azimuthal numbers. Herein we focus on the $LG_{0\pm1}$ modes, circled by dashed red squares in Fig. 1(g) and 1(h), which constitute the first excited manifold of the microcavity eigenstates.[24] For vector fields, the manifold is four-fold degenerate as the orbital angular momentum (OAM) $l$ and the spin angular momentum (SAM) $s$ both have a dimension of two, i.e., $l=\pm1$ representing $\pm2\pi$ phase winding and $s=\pm1$ representing right and left circular polarizations. Therefore, totally four eigenstates should be observed in the polarization-resolved spectrum if the degeneracy is fully lifted. However, this is never achieved in isotropic microcavities[28] even under the effect of TE-TM splitting.[24] To carry out a comprehensive characterization of the photonic properties of the confined microcavity, additional energy splitting can be introduced by the anisotropy of the active media. In the microcavity with circular confinement, the anisotropy leads to two different types of splitting: (1) the longitudinal one resulting in different resonant energies for orthogonally polarized states owing to the refractive index difference between the o and e lights, as already been demonstrated in Fig. 1; (2) the transverse one resulting in an effective ellipticity of the circular confinement potential, which further lift the degeneracy of the eigenstates of the same polarization. It should be noted that the physical mechanism of such transverse splitting is, however, not that straightforward. It is well known that the e-light has anisotropic distribution of refractive index which certainly leads to an effective elliptical deformation of the circularly-shaped potential. On the other hand, the o-light, in principle, should be completely isotropic and thereby no effective ellipticity can be induced by the refractive index distribution. Nevertheless, the TE-TM splitting of the cavity, instead of inducing vector vortex eigenstates with isotropic active media,[24] resulting in effective ellipticity of the confinement potential in a microcavity with anisotropic media. To clearly understand this mechanism, we start from the Hamiltonian of the open-access microcavity system:[23]

$$\hat{H} = \begin{pmatrix} \frac{\hbar^2 k^2}{2m} + \beta(k_x^2 - k_y^2) + \beta_0 + V + V'_c + V'_e & 2\beta k_x k_y \\ 2\beta k_x k_y & \frac{\hbar^2 k^2}{2m} - \beta(k_x^2 - k_y^2) - \beta_0 + V + V'_c \end{pmatrix} \quad (1)$$

where $m$ is the effective mass of the cavity photon or polariton, $k$ is the in-plane wavevector, $k_x$ and $k_y$ being the components of $k$ along the x and y direction, and $\varphi$ is the azimuthal angle. $\beta$ and $\beta_0$ represent the TE-TM splitting of the empty cavity and the energy splitting between orthogonally polarized states due to linear birefringence of α-perylene, respectively. The Hamiltonian is written in the linear polarization basis with $\begin{pmatrix}1\\0\end{pmatrix}$ and $\begin{pmatrix}0\\1\end{pmatrix}$ representing the polarizations along x (e-light) and y (o-light), respectively. Although some works reported the effect of emergent OA with perylene,[20,36] we don't include this term as our cavity does not satisfy the condition that the optical thickness has to make a half-wave plate, and moreover, the emergent OA has been proved to play no role in circularly confined systems.[37] The potential $V$ in Eq. (1) is the ideal circular potential formed due to the concave shape of the top cavity mirror which is expressed as $V = \frac{1}{2}m\omega^2(x^2 + y^2)$, where $\omega$ is the parameter characterizing the depth of the confinement potential. The term $V'_e$, which only apply to the e-light but not to the o-light, represents the elliptical deformation of the potential induced due to the anisotropy of refractive index, having the form $V'_e = \frac{1}{2}m\omega^2(x^2 - y^2)\delta_e$. Meanwhile, in reality the concave cavity mirror cannot be a perfectly symmetric circular shape, due to the imperfection of the FIB fabrication.[28] Therefore, we introduce in the model the elliptical asymmetry induced by the concave mirror $V'_c = \frac{1}{2}m\omega^2(x^2 - y^2)\delta_c$. It should be noted that $V'_c$ is not an intrinsic property of the physical problem but is induced mainly to simulate the experimental results. On the other hand, there is no intrinsic anisotropy for the o-light, which therefore does not introduce any potential corrections into the Hamiltonian.

To obtain the eigenvalues and eigenvectors of Eq. (1), we apply the degenerate perturbation theory in standard quantum mechanics,[38] by rewriting the Hamiltonian as:

$$\hat{H} = \hat{H}_0 + \hat{H}' = \begin{pmatrix} \dfrac{\hbar^2 k^2}{2m} + \dfrac{1}{2}m\omega^2(x^2+y^2) & 0 \\ 0 & \dfrac{\hbar^2 k^2}{2m} + \dfrac{1}{2}m\omega^2(x^2+y^2) \end{pmatrix}$$

$$+ \begin{pmatrix} \beta(k_x^2 - k_y^2) + \beta_0 + \dfrac{1}{2}m\omega^2(x^2-y^2)(\delta_c + \delta_e) & 2\beta k_x k_y \\ 2\beta k_x k_y & -\beta(k_x^2 - k_y^2) - \beta_0 + \dfrac{1}{2}m\omega^2(x^2-y^2)\delta_c \end{pmatrix} \quad (2)$$

in which $\hat{H}_0$ is the unperturbed Hamiltonian of a symmetric circular potential with known analytical solutions, and $\hat{H}'$ is the perturbation Hamiltonians describing the spin-orbit coupling terms and potential deformation terms, which are usually much weaker than the circular confinement potential and therefore can be treated as perturbations. To perform the perturbation theory, we use the Hermite-Gauss (HG) modes as the unperturbed eigenstates in the symmetric potential:

$$\psi_1 = \frac{xe^{-\frac{x^2+y^2}{2\sigma^2}}}{\sqrt{\pi/2\sigma^2}} \begin{pmatrix} 1 \\ 0 \end{pmatrix} \quad (3)$$

$$\psi_2 = \frac{xe^{-\frac{x^2+y^2}{2\sigma^2}}}{\sqrt{\pi/2\sigma^2}} \begin{pmatrix} 0 \\ 1 \end{pmatrix} \quad (4)$$

$$\psi_3 = \frac{ye^{-\frac{x^2+y^2}{2\sigma^2}}}{\sqrt{\pi/2\sigma^2}} \begin{pmatrix} 1 \\ 0 \end{pmatrix} \quad (5)$$

$$\psi_4 = \frac{ye^{-\frac{x^2+y^2}{2\sigma^2}}}{\sqrt{\pi/2\sigma^2}} \begin{pmatrix} 0 \\ 1 \end{pmatrix} \quad (6)$$

Where $\sigma = \sqrt{\hbar/m\omega}$ represent the mode dimension. Although the LG modes are the typical solutions under circular symmetry, HG modes can still be used as the unperturbed Hamiltonian as each LG function can be written as a linear superposition of HG functions. Meanwhile, the HG functions are better adapted to Cartesian coordinate which we have to choose to calculate the perturbation matrix in real space. According to the degenerate perturbation theory, each element of the perturbation matrix is calculated by:

$$H'_{mn} = (\psi_m, \hat{H}'\psi_n) \quad (7)$$

in which $m, n = 1, 2, 3, 4$. Herein the algorithm $k_x \rightarrow -i\dfrac{\partial}{\partial x}$, $k_y \rightarrow -i\dfrac{\partial}{\partial y}$ is applied, which yields the 4×4 perturbation matrix:

$$H' = \begin{pmatrix} \beta_0 + \dfrac{1}{2}(\delta_e + \delta_c)\hbar\omega + \dfrac{\beta}{\sigma^2} & 0 & 0 & \dfrac{\beta}{\sigma^2} \\ 0 & -\beta_0 + \dfrac{1}{2}\delta_c\hbar\omega - \dfrac{\beta}{\sigma^2} & \dfrac{\beta}{\sigma^2} & 0 \\ 0 & \dfrac{\beta}{\sigma^2} & \beta_0 - \dfrac{1}{2}(\delta_e + \delta_c)\hbar\omega - \dfrac{\beta}{\sigma^2} & 0 \\ \dfrac{\beta}{\sigma^2} & 0 & 0 & -\beta_0 - \dfrac{1}{2}\delta_c\hbar\omega + \dfrac{\beta}{\sigma^2} \end{pmatrix} \quad (8)$$

The eigenenergies and eigenstates of the system can then be determined by diagonalizing the perturbation matrix $H'$. The eigenvectors of $H'$ are the zeroth-order eigenfuctions of $\hat{H}$, and the eigenvalues are the corresponding first-order eigenenergy corrections of the system Hamiltonian $\hat{H}$.

Now we discuss on the perturbation matrix $H'$ in detail which is crucial to understand the underlying physics of our following experimental results. Let us suppose an ideal situation where the concave mirror is symmetric, i.e., $\delta_c = 0$, and first focus on the diagonal terms. First of all, the terms with $\pm\beta_0$ represent the energy splitting between orthogonally polarized states, i.e., the o and e lights, as appeared between $H'_{11}$ ($H'_{33}$) and $H'_{22}$ ($H'_{44}$) in Eq. (8), which is the major energy splitting in the system. Meanwhile, the term $\pm\delta_e$ induces a minor energy splitting between the states of the same polarization due to the anisotropy of the e-light, as seen between $H'_{11}$ and $H'_{33}$. Interestingly, a nontrivial term $\pm\beta/\sigma^2$ adds to the splitting between the same polarization, namely, between $H'_{11}$ ($H'_{22}$) and $H'_{33}$ ($H'_{44}$), constituting an effective ellipticity of the potential. While adding to the ellipticity caused by the anisotropy of the e-light ($H'_{11}$ and $H'_{33}$), it creates an emergent anisotropy for the o-light ($H'_{22}$ and $H'_{44}$) which is intrinsically isotropic. Such a phenomenon is enabled by the important underlying physical mechanism: although the TE-TM splitting is isotropic in all directions, it can induce anisotropic properties of the photonic eigenstates in an anisotropic environment, such as linear birefringence. It is in sharp contrast with the situations involving isotropic active media in which the TE-TM splitting leads to isotropic spin vortices.[24]

In addition to introducing emergent anisotropy, the term $\beta/\sigma^2$ appears as the off-diagonal elements of the perturbation matrix, introducing a coupling between the orthogonal polarized states. This term, however, is trivial as the absolute value of $\pm\beta_0$ is usually much larger than that of $\beta/\sigma^2$, which results in a large energy detuning between the coupled bare states (the corresponding diagonal elements) and almost completely eliminate the dressing effect. Otherwise, in the situation of isotropic material or weak linear birefringence ($\pm\beta_0 \approx 0$), the term $\beta/\sigma^2$ leads to a strong coupling between the o and e lights and form vector eigenstates of spin vortices.[24]

## IV. EXPERIMENTAL RESULTS AND DISCUSSION

The experiment fully demonstrating the anisotropic effects is carried out by measuring the microcavity emission with a grating of better spectral resolution (1200 grooves/mm), which exhibit fine spectral splittings of the $LG_{0\pm1}$ states in each polarization, labeled as x-i, ii and y-i, ii in Figures 2(a) and 2(b) for the e and o lights, respectively. A topographical imaging technique is applied (see Materials and Methods for detail) to obtain the real-space image of each cavity mode, as shown in Figs. 2(c) and 2(d). The modes display HG-like spatial distribution,[39-41] a typical effect of symmetry breaking induced by the effective ellipticity of the confinement potential. The e-light shows a larger fine energy splitting (0.866 meV or 0.192 nm) than the o-light (0.570 meV or 0.128 nm), which is expected as there is an additional effective ellipticity $\delta_e$ only for the e-light. Meanwhile, the spatial orientations of the modes for the e and o lights are also orthogonal, reflecting the fact that the long axis of the effective ellipticity of o and e lights are orthogonal to each other. This is reasonable considering the nature of TE-TM splitting: while propagating along x (resp. y) is TM (resp. TE) for the e-light, it is TE (resp. TM) for the o-light, and therefore the o and e lights represent orthogonal behaviors in all aspects: polarization, effective ellipticity and thereby the mode spatial orientation.

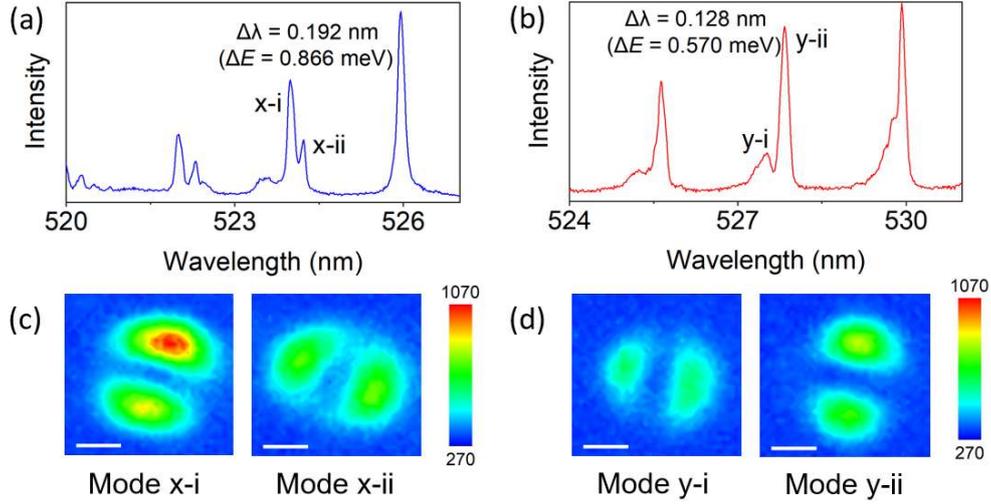

Figure 2: Measured μ-PL spectra and real-space images resolved in polarizaiton. (a) and (b) μ-PL spectra of the emission polarized in x (a) and y (b) directions, corresponding to the e and o light modes, respectively. The spectral separations $\Delta\lambda$ (energy separations $\Delta E$) between the fine split modes, i.e., x-i, ii and y-i, ii, are labeled. (c) and (d) The real space images of the modes x-i, ii (c) and y-i, ii (d) measured by tomography. Scale bar: 1 μm.

To further verify our understanding of the underlying physical mechanism, we rotate the planar bottom DBR together with the perylene microcrystals on it by 90°, while keeping the top concave DBR in its original orientation, and the measured spectra and topographical images are shown in Figure 3. Herein we label the modes formed by the e and o lights as x'-i, ii and y'-i, ii, respectively, whist the x' and y' axes are defined in the new reference frame after the rotation, which correspond to the same directions of y and x axes in the original reference frame of the lab, respectively. The key feature is that the orientations of the o-light (y'-i, ii) modes rotate 90° together with the optical axis of the perylene microcrystal, which clearly indicates that the o-light experiences anisotropy induced by the TE-TM splitting. Otherwise, if the o-light is isotropic, the orientation of modes would not have rotated as the actual ellipticity of the concave mirror is kept unchanged. We can easily know from the measured results that before the rotation, the effective ellipticity $\beta/\sigma^2$ adds to the concave mirror ellipticity $\delta_c$, and after the rotation, these two types of ellipticity partially cancels. That's why the corresponding fine energy splitting drops to 0.497 meV (0.112 nm), compared to 0.570 meV (0.128 nm) before the rotation, meaning that the circular potential becomes more symmetric, which makes the two split peaks of y'-i and y'-ii almost indistinguishable. It can also be known that the effect by $\beta/\sigma^2$ is stronger than that by $\delta_c$, otherwise the mode orientations would not have rotated. These conclusions apply well to the situation of the e-light modes, which in contrast show an increased fine splitting after the rotation (from 0.866 meV or 0.192 nm to 1.152 meV or 0.256 nm), indicating that $\delta_c$ partially cancels (resp. adds to) the combined effects of $\delta_e$ and $\beta/\sigma^2$ before (resp. after) rotating the microcrystal optical axis. The opposite trends of the change in the fine energy splitting between the o and e light modes with the 90° rotation reflect their nature of orthogonal polarizations.

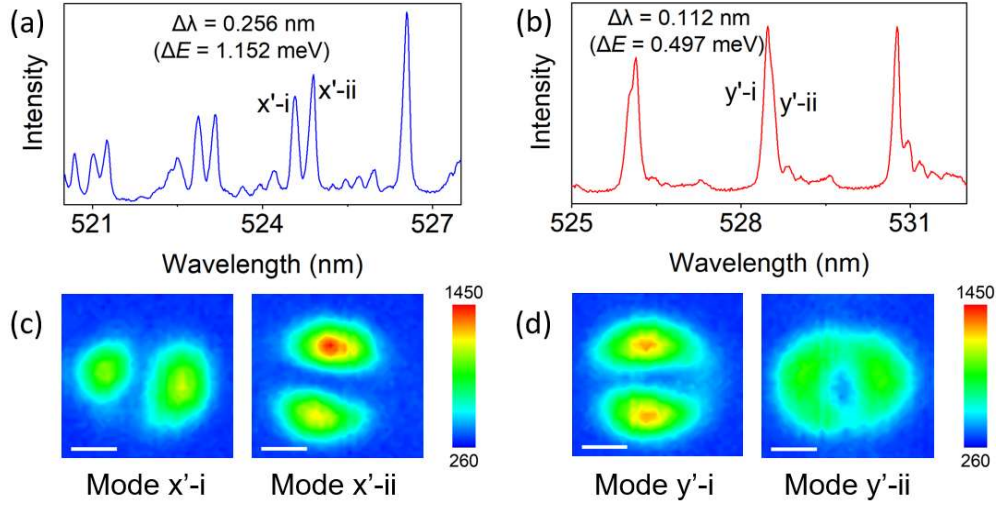

Figure 3: Measured μ-PL spectra and real-space images resolved in polarization, with the α-perylene microcrystal rotated by 90° with respect to the situation in Fig. 2. (a) and (b) μ-PL spectra of the emission polarized in x' (a) and y' (b) directions, corresponding to the e and o light modes, respectively. The spectral separations Δλ (energy separations ΔE) between the fine split modes, i.e., x'-i, ii and y'-i, ii, are labeled. Herein the x'-y' frame represents the reference frame after the 90° rotation, with the correspondence x'→y, y'→x to the x-y lab frame. (c) and (d) The real space images of the modes x'-i, ii (c) and y'-i, ii (d) measured by tomography. Scale bar: 1 μm. Note that the x'-y' frame is only used for labelling (or naming) the modes, while the graphs are still displayed in the x-y frame allowing better comparison with the situation before sample rotation.

Now that we know roughly the signs and quantitative relations of the relevant parameters, we can reproduce the experimental results by obtaining the eigenenergies and eigenstates from Eq. (8). The calculated eigenenergies and eigenfunctions are:

$$E_1^{(1)} = 8.860 \text{ meV} \tag{9}$$

$$E_2^{(1)} = -8.722 \text{ meV} \tag{10}$$

$$E_3^{(1)} = -8.087 \text{ meV} \tag{11}$$

$$E_4^{(1)} = 7.950 \text{ meV} \tag{12}$$

$$\Psi_1^{(0)} = 0.0164 \psi_2 - 0.9999 \psi_3 \tag{13}$$

$$\Psi_2^{(0)} = 0.0167 \psi_1 + 0.9999 \psi_4 \tag{14}$$

$$\Psi_3^{(0)} = -0.9999 \psi_2 - 0.0164 \psi_3 \tag{15}$$

$$\Psi_4^{(0)} = 0.9999 \psi_1 - 0.0167 \psi_4 \tag{16}$$

With the parameters are $\hbar\omega$ = 7.2 meV, $\sigma$ = 0.6 μm, $\beta$ = −0.1 meV·μm², $\beta_0$ = 8.4 meV, $\delta_c$ = 0.011 and $\delta_e$ = −0.06. The simulated energy splitting and spatial orientations of the cavity modes are presented in Fig. 4(a)-4(d), in good agreement with the experimental results in Fig. 2. Although the eigenfuctions have a mathematical form of linear superpositions of HG modes of orthogonal polarizations and spatial orientations, they can be approximately identified as pure linearly polarized HG modes as the amplitude differences of the two superposed HG components are extremely large. Otherwise, in the situation of near-zero $β_0$, the SO coupling would lead to equal amplitude in the superposition, resulting in spin vortices instead of anisotropy, as having been discussed before.

The theoretical reproduction of the situation after the 90° rotation is done in the x'-y' frame, in which the operation is equivalent to rotate the top concave mirror －90° while keeping the crystal not rotated. In this sense, the only difference to be introduced into Eq. (8) is to change the sign of $δ_c$, i.e., $δ_c = －0.011$, while keeping all the other parameters unchanged. Finally, the plotted spatial profiles and polarization orientations are rotated 90° back to the lab reference frame. The simulation results are shown in Figs. 4(e)-4(h), which reproduce well the experimental results in Fig. 3.

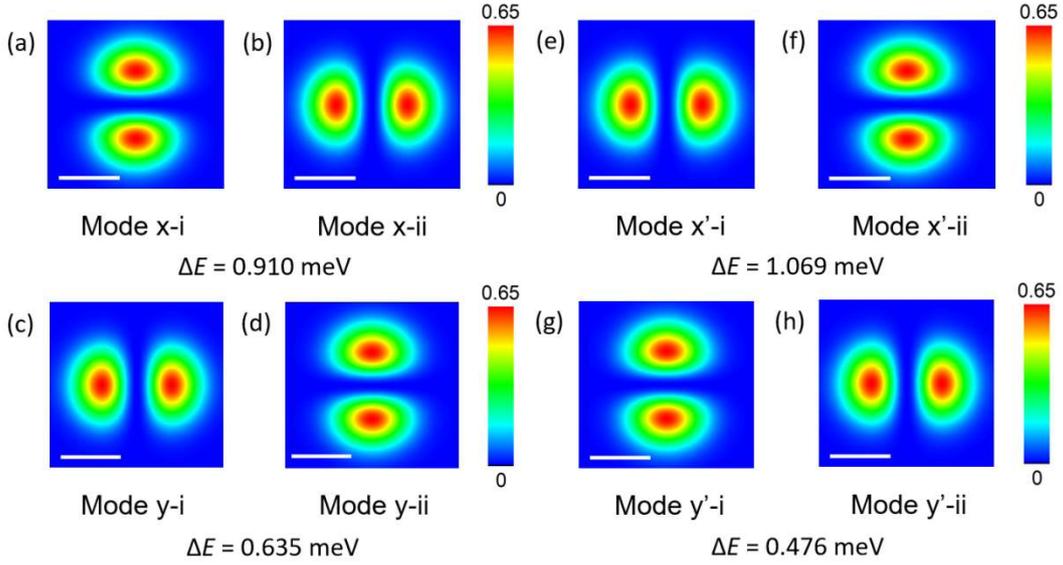

Figure 4: The simulated mode spatial profiles. (a-d) before the rotation, (e-h) after the rotation. Scale bar: 1 μm. The fine energy splitting is written under each graph.

## V. CONCLUSION

In conclusion, we demonstrated that the SO coupling induced by TE-TM splitting introduces emergent anisotropy in a laterally confined microcavity embedded with anisotropic active media, even if the TE-TM splitting, the cavity confinement and the light propagation are all intrinsically isotropic. This mechanism, which is well explained theoretically using the degenerate perturbation theory, reveals the unique role of TE-TM splitting in anisotropic cavity media and thereby significantly extend the area of microcavity spinoptronics. Following this mechanism, a flexible engineering of the cavity mode features, in both spatial distribution and polarization, can be reached via the interplay among various parameters including the TE-TM splitting, material anisotropy and cavity shape deformation, etc. In addition, this mechanism can be potentially applied for determining material properties, such as the anisotropic feature and refractive indices in certain emerging materials which would be otherwise difficult to know via conventional measurements, by analyzing the photonic modes of the embedded microcavity.

**Materials and Methods**

**Synthesis of the organic microcrystals.** The α-perylene was prepared by the space-confined self-assembly method.[20,31] First, perylene was added into the chlorobenzene to get the 0.5mg/mL perylene/chlorobenzene solution. 50μL solution was drew with microliter syringe and added into the 1mg/mL TBAB/deionized water solution. When the chlorobenzene was fully evaporated, perylene was left on the solution surface and formed the crystalline α-perylene with two-dimensional square shape. The microcrystals were transferred onto the bottom DBR by simply contacting the DBR and the surface of liquid.

**Configuration of the open-access microcavity.** The top concave mirror was fabricated by first making arrays of concave shapes on a silica substrate by focused ion beam (FIB) milling, which exhibit radii of curvature (ROC) of 50 μm, 20 μm, 12 μm and 7 μm. In this work the concave shape has ROC=20 μm. Afterwards, 9 pairs of $Ta_2O_5/SiO_2$ bilayers were deposited on the substrate to form a DBR with a center wavelength of 490 nm and a broad reflectivity bandwidth of ~108 nm. The bottom mirror was fabricated by depositing 9 pairs of $Ta_2O_5/SiO_2$ layers to form a DBR with a center wavelength of 540 nm and a broad reflectivity bandwidth of ~80 nm. The top and bottom mirrors are controlled individually by two groups of Attocube nanopositioners.

**The experiment setup for polarization resolved μ-PL.** The real-space spectroscopy is performed at room temperature. A continuous laser with a wavelength of 405 nm is focused by a 100× objective lens (NA=0.8) to excite the sample. The light emitted by the sample is collected by the same objective lens and imaged in front of the entrance slit of the spectrometer (Andor SR750). The light is dispersed by the spectrometer and detected by an EMCCD (Andor DU970P-BVF). The polarization-resolved measurements were done by inserting a half-wave plate and a linear polarizer in front of the spectrometer entrance slit. To take topographical images, we move the lens in front of the spectrometer entrance slit successively with each small step of 0.1μm, to obtain the spectra resolved in y-direction, while each spectrum corresponding to a specific position along x-direction. Finally, the real-space images are reconstructed by jointing all the y-resolved images at the specific wavelength of the optical modes in the order of the position along x.

## ACKNOWLEDGEMENT

This research was funded by the National Natural Science Foundation of China (12074303, 11804267, and 22150005), Shaanxi Key Science and Technology Innovation Team Project (2021TD-56), Natural Science Foundation of Beijing, China (KZ202110028043).

## AUTHOR DECLARATIONS

### Conflict of Interest

The authors declare no conflict of interest.

## DATA AVAILABILITY

The data that support the findings of this study are available from the corresponding author upon reasonable request.